\documentclass[twocolumn]{aastex631}
\usepackage{textcomp}
\usepackage{gensymb}
\usepackage{amsmath}
\begin{document}

\title{A Target Search for Fast Radio Bursts Associated with Two Fast Blue Optical Transients: AT2018cow and CSS161010}

\correspondingauthor{Xiang-Dong Li}
\email{lixd@nju.edu.cn}

\author[0000-0002-0822-0337]{Shi-Jie Gao}
\author[0000-0002-0584-8145]{Xiang-Dong Li}
\affiliation{School of Astronomy and Space Science, Nanjing University, Nanjing, 210023, People's Republic of China}
\affiliation{Key Laboratory of Modern Astronomy and Astrophysics, Nanjing University, Ministry of Education, Nanjing, 210023, People's Republic of China}
\author[0000-0001-5684-0103]{Yi-Xuan Shao}
\author[0000-0002-5683-822X]{Ping Zhou}
\affiliation{School of Astronomy and Space Science, Nanjing University, Nanjing, 210023, People's Republic of China}
\affiliation{Key Laboratory of Modern Astronomy and Astrophysics, Nanjing University, Ministry of Education, Nanjing, 210023, People's Republic of China}
\author[0000-0002-3386-7159]{Pei Wang}
\affiliation{National Astronomical Observatories, Chinese Academy of Sciences, Beijing, People's Republic of China}
\affiliation{Institute for Frontiers in Astronomy and Astrophysics, Beijing Normal University, Beijing 102206, People's Republic of China}
\author[0000-0002-1067-1911]{Yun-Wei Yu}
\affiliation{Institute of Astrophysics, Central China Normal University, Wuhan 430079, People's Republic of China}
\affiliation{Key Laboratory of Quark and Lepton Physics (Ministry of Education), Central China Normal University, Wuhan 430079, People's Republic of China}
\author[0000-0002-9322-9319]{Zhen Yan}
\affiliation{Shanghai Astronomical Observatory, Chinese Academy of Sciences, Shanghai 200030, People's Republic of China}
\affiliation{School of Astronomy and Space Science, University of Chinese Academy of Sciences, Beijing 100049, People's Republic of China}
\author[0000-0003-3010-7661]{Di Li}
\affiliation{New Cornerstone Science Laboratory, Department of Astronomy, Tsinghua University, Beijing 100084, China}
\affiliation{National Astronomical Observatories, Chinese Academy of Sciences, Beijing, People's Republic of China}

\begin{abstract}
Fast blue optical transients (FBOTs) are luminous, rapidly evolving events with blue spectra, possibly powered by newborn magnetars and linked to fast radio bursts (FRBs). Given this potential connection, we conducted deep radio observations of two nearby FBOTs (AT2018cow and CSS161010) using the Five-hundred-meter Aperture Spherical radio Telescope (FAST), but detected no FRB-like signals. Our observations establish the most stringent upper limits on millisecond radio transients from FBOTs, reaching $\sim$10 mJy flux density. Assuming a log-normal luminosity function analogous to the repeating FRB 121102, we constrain the burst rate from potential magnetars in FBOTs to $<0.01$ hr$^{-1}$. The short ejecta escape timescale ($\sim$2.6 yr) compared to our observation epochs (4$-$6 years post-explosion) suggests that nondetection may not be attributed to FBOT's ejecta absorption. These findings impose useful constraints on the FRB activity emanating from newborn magnetars within FBOTs. They indicate that if there is a burst phase, it is either characterized by weaker bursts, occurs less frequently compared to those in known repeating FRB sources, or takes place beyond the time frame of our current observations.
To gain deeper insights into the birth-related activity of magnetars, it is of importance to conduct timely and sustained FRB searches in FBOTs that emerge in the future.

\end{abstract}


\section{Introduction} \label{sec:intro}

Fast Radio Bursts (FRBs) are millisecond-duration, high-energy and extremely bright radio transients predominantly originating from cosmological distances \citep{Lorimer+2007,Thornton+2013,Petroff+2019R,Cordes+2019,Zhang+2020,Petroff+2022R,Zhang+2023R}. They occur uniformly across the sky at a rate of a few thousand per day \citep{Lorime+2024}. As of now, 865 FRB sources have been discovered\footnote{\url{https://blinkverse.alkaidos.cn/}} \citep{Xu+2023}, with 71 of them being repeaters. Notably, 96 of these FRB sources have been localized to their respective host galaxies, playing a crucial role in cosmological research \citep{Gordon+2023,Michilli+2023,Law2024,Bhardwaj+2024,Sharma+2024,Sherman+2024,Shannon+2025,Wang+2025,chimefrb+2025,Connor+2025}.

The origin of FRBs remains a topic of intense research. While multiple progenitor scenarios have been proposed, mounting evidence connects FRBs to magnetar activity. Various magnetar-based models have been proposed, suggesting that FRBs may arise from processes such as magnetar flares \citep{Popov+2010,Kulkarni+2014,Lyubarsky+2014}, repeating bursts from millisecond magnetars \citep{Metzger+2017,Beloborodov+2017,Margalit+2019,Metzger+2019}, pair cascade driven magnetic field twists \citep{Wadiasingh+2019}, enhanced magnetic reconnection in magnetar current sheets \citep{Lyubarsky+2020}, coherent inverse Compton scattering \citep{Zhang+2022}, free-electron laser mechanism \citep{Lyutikov+2021} and magnetized shocks \citep{Thompson+2023}. The detection of FRB~200428, associated with the Milky Way magnetar SGR~J1935+2154 confirmed that at least some FRBs originate from processes linked to magnetars \citep{Bochenek+2020,ChimeFrbJ1935+2020,Mereghetti+2020,Li+2021NatAs,Ridnaia+2021NatAs,Tavani+2021NatAs,Borghese+2022}.

Fast blue optical transients (FBOTs) represent another class of extreme transients characterized by 
\begin{itemize}
\item rapid photometric evolution (with durations above half-maximum brightness lasting $\lesssim12~{\rm d}$),
\item high optical luminosities (up to $\gtrsim 10^{44}~{\rm erg~s^{-1}}$), 
\item blue colors  \citep[with effective temperature $T_{\rm eff}\sim 2\times10^4~{\rm K}$,][]{Drout+2014,Pursiainen+2018,Inserra+2019}.
\end{itemize}
Since the discovery of the prototype FBOT AT2018cow \citep{Prentice+2018}, over 100 FBOTs have been discovered \citep{Ho+2023}, including AT2018lug \citep{Ho+2020+at2018lug}, CSS161010 \citep{Coppejans+2020}, AT2020xnd \citep{Perley+2021}, AT2020mrf \citep{Yao+2021} and AT2023fhn \citep{Chrimes+2024}.

The nature of FBOTs also remains uncertain. The proposed models are classified into central engine models and shock interaction models \citep{Liu+2023}. In the former models, FBOTs are thought to arise from low-mass ejecta powered by a central engine, which could be a rapidly rotating magnetar \citep{Kasen+2010,Inserra+2013,Drout+2014,Pursiainen+2018,Prentice+2018,Li+2024}, or a newly formed accreting stellar-mass black hole \citep{Kashiyama+2015,Yao+2021}. Their fast evolution is attributed to an ejecta mass $\sim 0.1\,{\rm M_{\odot}}$, and luminous brightness to additional energy injection from a central engine rather than the radioactive decay of $^{56}\rm Ni$ \citep[e.g.,][]{Drout+2014,Pursiainen+2018}. Systems composed of low-mass ejecta and a magnetar may originate from ultra-stripped supernovae \citep{Tauris+2013,Tauris+2015,Tauris+2017}, electron-capture supernovae \citep{Moriya+2016,Mor+2023}, accretion-induced collapse of white dwarfs \citep{Wangbo+2020}, or merges of double white dwarfs, double neutron stars or neutron star--white dwarf binaries \citep{Lyutikov+2019,Yu+2019,Zenati+2019,Lyytikov+2022}.

Compared with other known magnetars, those potentially associated with FBOTs, if they indeed exist, are likely among the youngest and may exhibit more magnetospheric activities to trigger FRBs. These newborn NS may rotate rapidly and deferentially, allowing their magnetic fields to be significantly amplified through magnetorotational instability \citep{Duncan+1992,Dessart+2007,Cheng+2014}. Compared to other transients likely powered by magnetars, such as long gamma-ray bursts and super-luminous supernovae, FBOTs tend to have lower ejecta masses \citep{Liu+2022}, which makes it easier for potential radio bursts to escape. These factors motivate the observation of FBOTs in search for FRBs, using the high-sensitivity and high-time-resolution data from the Five-hundred meter Aperture Spherical radio Telescope \citep[FAST,][]{Nan+2008,Nan+2011}. 

In 2022, We used FAST to observe two nearby FBOTs, AT2018cow and CSS161010, in search of FRB-like signals. We report our results in this paper. The rest of this paper is organized as follows: \autoref{sec:obs} details our FRB search observations and data analysis techniques, \autoref{sec:res} presents the search results, and discussion and we conclude in \autoref{sec:con}. In this paper, we employ the $\Lambda$CDM model for distance measures in cosmology and use the parameters from the latest cosmic microwave background radiation measurements, i.e., $H_0=67.7~{\rm km~s^{-1}~Mpc^{-1}}$, $\Omega_{\rm M}=0.31$, $\Omega_{\Lambda}=0.69$ \citep{Planck+2020}. 

\section{Observations and Data Reduction}\label{sec:obs}

\begin{figure*}
    \centering
    \includegraphics[width=\linewidth]{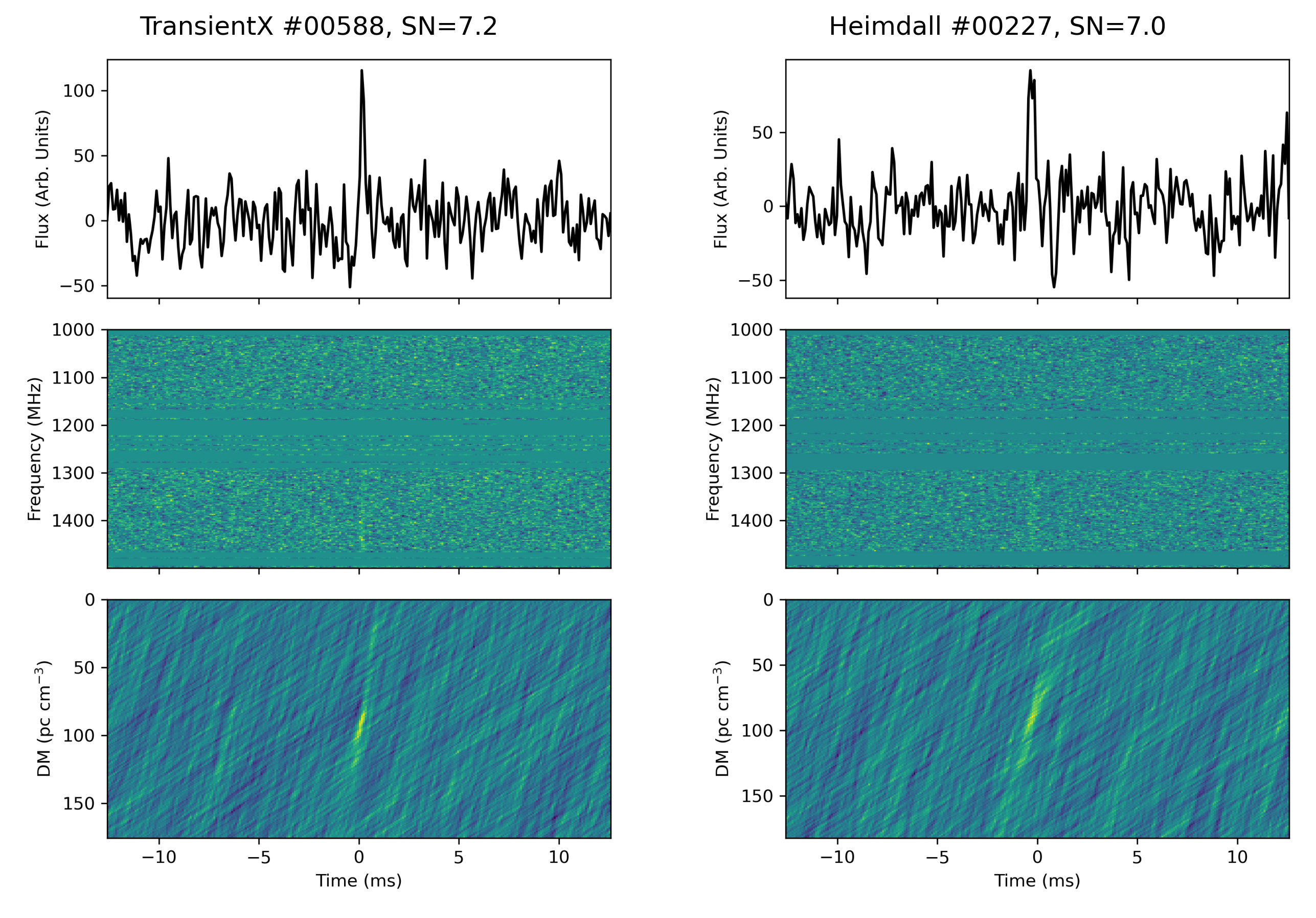}
    \caption{Two pulses from RRAT~J0628+09 used to test search pipelines. The left and right columns show two low-S/N candidates detected with \texttt{TransientX} and \texttt{Heimdall}, respectively. In each column, the top panels show dedispersed flux versus time, the middle panels show intensity versus dedispersed frequency and time, and the bottom panels show DM versus time. The search pipelines yield the best DM values of $~89$ and $~91~{\rm pc~cm^{-3}}$ for the pulses in the left and right column, respectively.\label{fig:testpipeline}}
\end{figure*}

\begin{figure*}
    \centering
    \includegraphics[width=\linewidth]{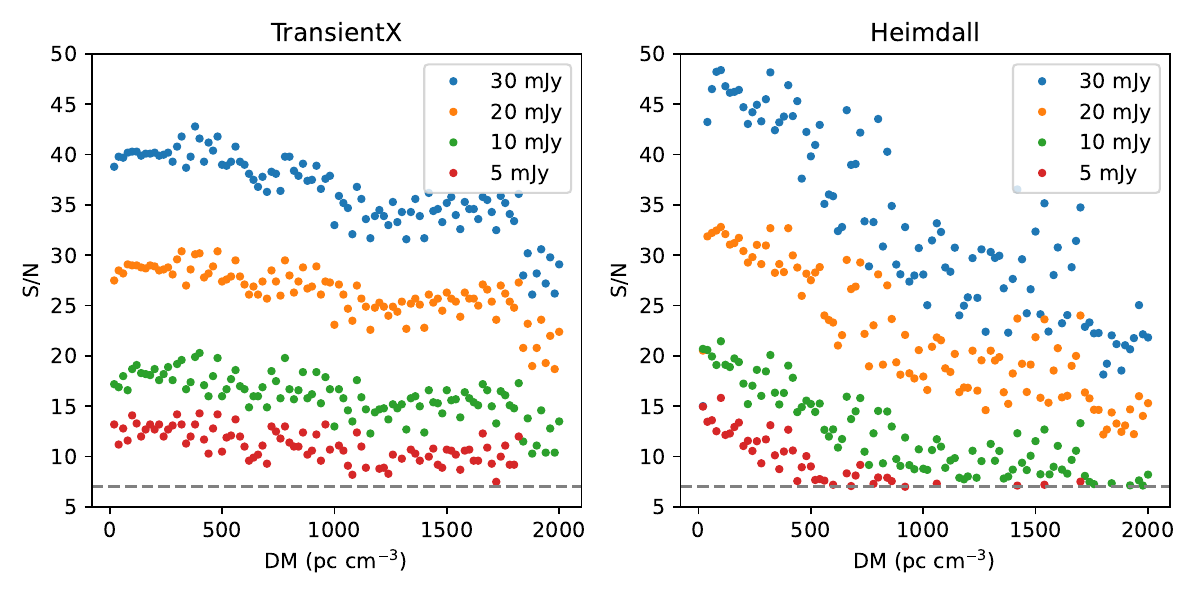}
    \caption{Detection completeness of the single-pulse search pipeline assessed via synthetic pulse injection using \texttt{FRB Fake}. Injected flux densities are 5 (red), 10 (green), 20 (orange) and 30 (blue) mJy. The detected S/N are plotted against injected DMs for \texttt{TransientX} (left panel) and \texttt{Heimdall} (right panel). S/N decreases at higher DMs due to increased DM smearing. \label{fig:fake}}
\end{figure*}

\subsection{Observation Setup}
Considering the luminosity distances $d_{\rm L}$ (or redshift $z$) and positions of known FBOTs, we selected two FBOTs AT2018cow and CSS161010 as our targets. The prototype FBOT AT2018cow is located at $\alpha_{\rm J2000}=16{\rm ^h}16{\rm ^m}00{\rm ^s}.25$, $\delta_{\rm J2000}=+22\degree16'05''.1$, with a redshift of $z=0.014$ and a luminosity distance of $d_{\rm L}=61~{\rm Mpc}$ \citep{Prentice+2018}. CSS161010 is located at $\alpha_{\rm J2000}=04{\rm ^h}58{\rm ^m}34{\rm^s}.00$, $\delta_{\rm J2000}=-08\degree18'03''.0$, with a redshift of $z=0.034$ and a luminosity distance of  $d_{\rm L}=150~{\rm Mpc}$ \citep{Coppejans+2020}.

AT2018cow was observed three times on August, October and December 2022, and each observation lasted 1 hour; CSS161010 was observed twice on September and December 2022 for 45 minutes and 1 hour, respectively\footnote{Project ID: PT2022\_0143, PI: Xiang-Dong Li.}. Observations were carried out with the center beam of FAST's $L$-band 19-beam receiver \citep{LiDi+2018} at a central frequency of 1250 MHz using a bandwidth of 500 MHz. The beam size (half-power beamwidth) is $3'$ \citep{LiDi+2018,Jiang+2020}, which has a good coverage of the whole host galaxies for the two FBOTs. Data were accumulated to a time resolution of $49.152~{\rm \mu s}$ and then written to PSRFITS-format \citep{psrfits} files with 4096 frequency channels and four polarizations. To calibrate polarizations for each observation, we turned on the noise diode in a 0.2~s cycle for 1 minute at both the beginning and the end of each observation.

\subsection{Single-pulse Search Pipeline}

We used \texttt{PulsarX}'s \citep{PulsarX} \texttt{filtool} command to mitigate radio frequency interference (RFI), optimize the bandpass and convert PSRFITS-format to filterbank-format \citep{filterbank} data with 1024 frequency channels for subsequent single-pulse searches. Single-pulse searches were conducted using both \texttt{TransientX}\footnote{\url{https://github.com/ypmen/TransientX}} \citep{TransientX} and \texttt{Heimdall}\footnote{\url{https://sourceforge.net/projects/Heimdall-astro}} \citep{Barsdell+2024} to perform a double cross check. \texttt{TransientX} is a high-performance CPU-based single-pulse search software, and \texttt{Heimdall} is a GPU-accelerated pipeline for radio transient detection. 

Based on the electron density model YWM16 \citep{YMW16}, the total dispersion measure (DM) contributions from the Milky Way and the intergalactic medium are $201$ and $207~{\rm pc~cm^{-3}}$ for AT2018cow and CSS161010, respectively. A trial DM range up to $2000~{\rm pc~cm^{-3}}$ was used to account for unknown contributions from their host environments. For pulse searches with \texttt{TransientX}, a dedispersion grid with a DM range of $5-2000~{\rm pc~cm^{-3}}$ and step size of $0.1-1~{\rm pc~cm^{-3}}$ was generated by using the \texttt{PRESTO}'s Python script \texttt{DDplan.py} \citep{Ransom+2011}. The maximum pulse width was set to the default value of $\sim 50~{\rm ms}$ and a signal-to-noise ratio (S/N) threshold of 7.0 was applied. In \texttt{Heimdall}, a similar DM range of $5-2000~{\rm pc~cm^{-3}}$ was used, with a dedispersion plan implemented within the software. The maximum boxcar width was set to 1024 samples, corresponding to a maximum pulse width of $\sim 50~{\rm ms}$. Candidates with similar boxcar or DM trials were clustered within \texttt{Heimdall} and a S/N threshold of 7.0 was applied.

Diagnostic plots, including the dedispersed pulse, dedispersed frequency versus time, and pulse search in DM space, were generated by using the single-pulse analysis software \texttt{YOUR} \citep{Aggarwal+2020} for  candidates from both pipelines. We employed \texttt{FETCH} \citep{FETCH}, an open-source machine learning tool utilizing convolutional neural networks to analyze information from the diagnostic plots. \texttt{FETCH} offers a variety of trained models, and a candidate is considered promising if at least one model classifies it as favorable. Filtered plots were then visually inspected to identify the most promising single-pulse candidates.

\subsection{Pipeline Validation}

Before reducing the data from the two FBOTs, we tested our single-pulse search pipeline using available data from the rotating radio transient RRAT~J0628+09, observed with the same parameters\footnote{Project ID: PT2021\_0117, PI: Xiang-Dong Li.}. RRAT~J0628+09 is a Galactic, extremely sporadic radio emitter with a spin period of 1.2~s and a ${\rm DM} \simeq 88~{\rm pc~cm^{-3}}$ \citep{Cordes+2006}. The left and right panels of \autoref{fig:testpipeline} display two low-S/N candidates for RRAT~J0628+09 discovered with \texttt{Heimdall} and \texttt{TransientX}, respectively. The plots show the dedispersed pulse diagram, dynamical spectrum and pulse search in time and DM space from top to bottom. The successful detection of faint radio pulses confirms that our pipelines work as expected.

To quantitatively evaluate the detection completeness of our pipeline across varying flux densities and DM, we  employed the \texttt{FRB~Fake}\footnote{Author: Leon Houben. Code available at \url{https://gitlab.com/houben.ljm/frb-faker}.} tool to inject synthetic single pulses into the filterbank data obtained from the observations of AT2018cow. The injected pulses were modeled as Gaussian profiles with a full width at half maximum of 1 ms. We injected pulses with flux densities ($F_{\rm in}$) of $5$, $10$, $20$, and $30{\rm ~mJy}$, assuming a fixed system equivalent flux density ($T_{\rm sys}/G$) of 1.25{\rm ~Jy} for the FAST $L$-band 19-beam receiver \citep{Jiang+2020}. Here $T_{\rm sys}$ is the system temperature, and $G$ is the telescope gain. For each $F_{\rm in}$, we sampled 100 DM values ranging from $20$ to $2000~{\rm pc~cm^{-3}}$ in steps of $20~{\rm pc~cm^{-3}}$. The simulated data were then processed through our single-pulse search pipeline.

The results are shown in \autoref{fig:fake}, where the detected S/N is plotted against the injected DM for \texttt{TransientX} (left panel) and \texttt{Heimdall} (right panel), respectively. In the \texttt{TransientX} pipeline, no pulses were missed when $F_{\rm in}\ge 10~{\rm mJy}$, and only 15 pulses were missed at ${\rm DM}\gtrsim 1000~{\rm pc~cm^{-3}}$ for $F_{\rm in}=5~{\rm mJy}$. In the \texttt{Heimdall} pipeline, no pulses were missed when $F_{\rm in}\ge20~{\rm mJy}$, only 8 pulses were missed at $F_{\rm in}=10~{\rm mJy}$ and ${\rm DM}\gtrsim1250~{\rm pc~cm^{-3}}$, and 48 pulses were missed at $F_{\rm in}=5~{\rm mJy}$ and ${\rm DM}\gtrsim550~{\rm pc~cm^{-3}}$.
The missed pulses were primarily due to the reduced S/N caused by DM smearing at higher DM values, which also explains the decreasing trend seen in the DM--S/N plots. The S/N values from \texttt{TransientX} decrease more gradually with increasing DM compared to \texttt{Heimdall}, owing to its finer dedispersion grid and less aggressive downsampling. For our two FBOT targets, the DM contributions from the Milky Way and the intergalactic medium are approximately $200~{\rm pc~cm^{-3}}$, well below the threshold where detection incompleteness becomes significant. In summary, our pipeline, utilizing  both \texttt{TransientX} and \texttt{Heimdall} for cross-verification, achieves complete detection for pulses with flux densities as low as $5-10~{\rm mJy}$.

\section{Results and Discussion}\label{sec:res}
\begin{figure*}[hb]
    \centering
\includegraphics[width=\linewidth]{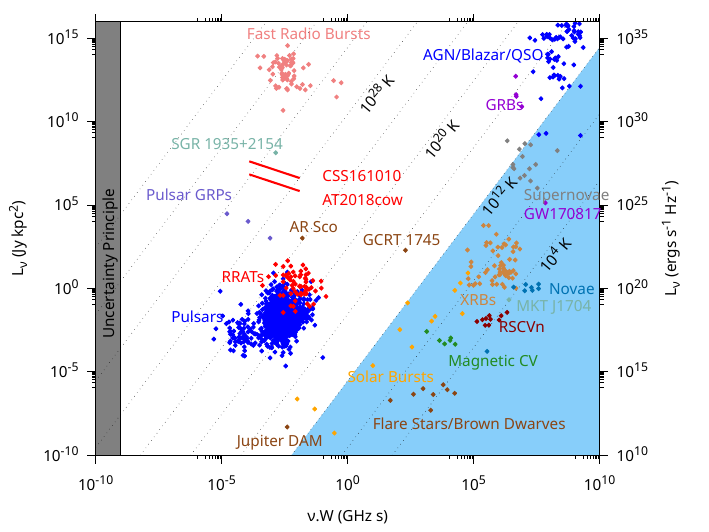}
    \caption{Currently known radio sources in the space of the pseudo-luminosity and the products of the observing frequency and transient/variability timescale. The red lines indicate the pseudo-luminosity upper limits for AT2018cow and CSS161010. The assumed pulse width $W$ for the two sources is in the range of $0.001-0.1~{\rm s}$ and the observing frequency $\nu$ is taken as $1.25~{\rm GHz}$. Data and figure courtesy of Manisha Caleb, adapted from \citet{Hurley-Walker+2022}, originally from \citet{Pietka+2015}. \label{fig:transient}}
\end{figure*}

Single-pulse searches for AT2018cow and CSS161010 produced 6827 candidates. After classification by \texttt{FETCH} and filtering, the candidates were visually inspected and identified as RFI. The number 6827 is small enough to allow manual inspection, so we checked each candidate individually. No candidate was found to have an astrophysical origin, which confirms the accuracy of \texttt{FETCH}. Consequently, we conclude that no significant single-pulse candidates were detected from AT2018cow and CSS161010.

We next discuss the implications of these nondetections, including search sensitivity, the luminosity function and burst rate of FRBs from magnetars, and the time-scales for FRB's radiation to escape the FBOT ejecta.

\subsection{Search Sensitivities}
\begin{table*}
    \setlength{\tabcolsep}{5pt}
    \centering
        \caption{The source name, start time, observation length, zenith angle (ZA), system temperature $T_{\rm sys}$ , zenith-sensitive telescope gain $G$ and the derived flux upper limit $S_{\rm min}$ for each search observation. The second and third parts show the results from the calibration observation carried out at the begining and the end of each search observation, respectively.\label{tab:smin}}
    \begin{tabular}{ccc|cccc|cccc}
\hline
&&&\multicolumn{4}{c|}{calibration 1}&\multicolumn{4}{c}{calibration 2}\\
Name & Start Time &Length& ZA&$T_{\rm sys}$ & $G$ &$S_{\rm min}$&ZA &$T_{\rm sys}$& $G$ &$S_{\rm min}$\\
&(MJD)&(s)&$(\deg)$&(K)&$(\rm K~Jy^{-1})$&$(\rm mJy)$&$(\deg)$&(K)&$(\rm K~Jy^{-1})$&$(\rm mJy)$\\
\hline
AT2018cow &59822.3823 & $3600$ &  18.7 & $ 23.9\pm  0.3$ & $ 16.2\pm 0.12$ & $ 13.1\pm  0.2$ &  6.0 & $ 20.5\pm  0.1$ & $ 16.0\pm  0.1$ & $ 11.4\pm  0.1$ \\
AT2018cow &59853.2587 & $3600$ &  31.3 & $ 23.2\pm  0.6$ & $ 14.5\pm 0.16$ & $ 14.2\pm  0.4$ & 18.0 & $ 18.8\pm  0.5$ & $ 16.2\pm  0.1$ & $ 10.3\pm  0.3$ \\
AT2018cow &59886.2122 & $3600$ &  17.2 & $ 20.1\pm  0.5$ & $ 16.1\pm 0.12$ & $ 11.1\pm  0.3$ &  4.8 & $ 22.7\pm  0.1$ & $ 16.0\pm  0.1$ & $ 12.6\pm  0.1$ \\
CSS161010 &59824.9330 & $2700$ &  35.3 & $ 25.4\pm  0.6$ & $ 13.0\pm 0.17$ & $ 17.3\pm  0.5$ & 33.9 & $ 28.9\pm  0.5$ & $ 13.5\pm  0.2$ & $ 19.0\pm  0.4$ \\
CSS161010 &59889.7539 & $3600$ &  35.5 & $ 23.1\pm  0.3$ & $ 13.0\pm 0.17$ & $ 15.9\pm  0.3$ & 34.2 & $ 26.5\pm  0.2$ & $ 13.4\pm  0.2$ & $ 17.5\pm  0.3$ \\
\hline
    \end{tabular}
\end{table*}

The sensitivity limit of our transient searches can be derived using the radiometer equation
\begin{equation}
    S_{\rm min}={\rm S/N} \times \frac{\beta T_{\rm sys}}{G\sqrt{N_{\rm p}\Delta\nu W}},
\end{equation}
\citep[e.g.,][]{McLaughlin+2003}. Here, $\beta=1.1$ is a correction factor for digital losses, $N_{\rm p}=2$ is the number of summed polarizations, $\Delta\nu$ is the effective bandwidth, and $W$ is the observed width of the pulse. We extracted the calibration data and processed it using \texttt{DSPSR} \citep{vanStraten+2011}. The system temperature ($T_{\rm sys}$) during a specific observation can be calculated as follows \citep{pulsarBook},
\begin{equation}
    \frac{T_{\rm sys}}{T_{\rm cal}}=\frac{\rm OFF}{\rm OFFCAL-OFF},
\end{equation}
where $T_{\rm cal}=12.1~{\rm K}$ is the equivalent temperature of the noise diode for FAST\footnote{\url{https://fast.bao.ac.cn/cms/article/193/}}, OFFCAL and OFF represent the counts of system noise with and without the noise diode, respectively. Using \texttt{PSRCHIVE} \citep{Hotan+2004}, we obtained the values of OFFCAL and OFF values and their uncertainties. No flux calibration observation on standard candles was conducted, so we used the zenith-sensitive gain factor by adopting fitted formula in \citet{Jiang+2020}. The observed single-pulse width was assumed to be $W=1~{\rm ms}$, with broadening effects from smearing and scattering taken into account. The effective bandwidth $\Delta \nu$ is about 75\% of the total bandwidth ($500~{\rm MHz}$) after removing RFI and bad edges. Therefore, we derived the flux sensitive limit for our transient searches, with the results detailed in \autoref{tab:smin}. The second and third parts present the calibration results for observations at the beginning and end of each session, respectively. The system temperatures are in the range of $20-29~{\rm K}$, and the telescope gains are in the range of $11-16~{\rm K~Jy^{-1}}$. These parameters are consistent with the expected performance of FAST \citep{Jiang+2020}. The upper limit of the flux for single-pulse searches targeting FBOTs can reach as low as $\sim 10~{\rm mJy}$, which represents the most stringent constraint on the radio fluxes of millisecond-level radio signals from FBOTs.

\autoref{fig:transient} shows the known radio transients in the plane of the pseudo-luminosity ($L_\nu$) and the products of the frequency ($\nu$) and the transient/variability timescale ($W$). The red dashed lines show the pseudo-luminosity upper limits for AT2018cow and CSS161010 assuming $W\sim 1-100\,{\rm ms}$ with fixed $T_{\rm sys}=24\,{\rm K}$ and $G=16~{\rm K~Jy^{-1}}$. Our searches were sensitive to FRB-like bursts similarly to those from the Galactic magnetar SGR~J1935+2154.

\subsection{Luminosity Function of FRBs}
\begin{figure*}
    \centering
    \includegraphics[width=0.8\linewidth]{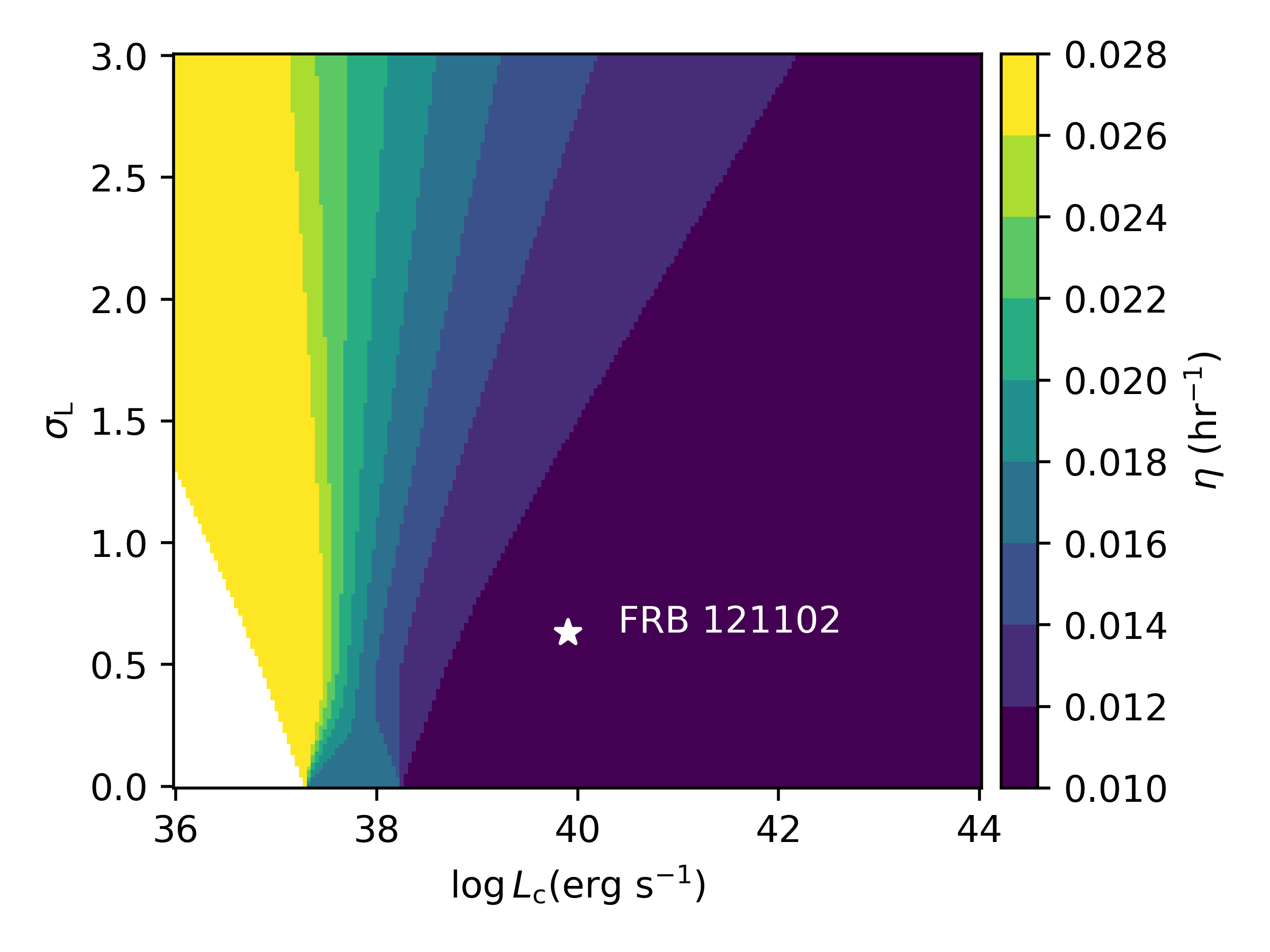}
    \caption{The plot of the burst rate ($\eta$) versus $L_{\rm c}$ and $\sigma_{\rm L}$ for a log-normal luminosity function, assuming a total nondetection probability of $P_{\rm non}=95\%$. The color in each pixel corresponds to  $\eta$ required for given $L_{\rm c}$ and $\sigma_{\rm L}$ to remain consistent with our non-detection. The values of $\sigma_{\rm L}$ and $L_{\rm c}$ for FRB~121102 \citep{Men+2019} are indicated with a white star.}
    \label{fig:lf}
\end{figure*}

Based on our nondetections, we attempt to test the hypothesis of association between magnetars and FRBs, and constrain the burst rate of repeating FRBs from magnetars. The luminosity function of repeating FRBs can be assumed to a logarithmic normal distribution \citep[e.g.,][]{BurkeSpolaor+2012,Levin+2012,Gourdji+2019},
\begin{equation}
    \phi(L|L_{\rm c},\sigma_{\rm L})=\frac{1}{\sqrt{2\pi}\sigma_{\rm L}L}\exp\left[-\frac{(\log L-\log L_{\rm c})^2}{2\sigma_{\rm L}^2}\right],
\end{equation}
where $\log L_{\rm c}$ and $\sigma_{\rm L}$ are the mean  and standard deviation of $\log L$, respectively. We assume that the number $k$ of radio pulses from potential young magnetars in FBOTs follow the same Poisson's distribution
\begin{equation}
    P(k|T_{{\rm obs},i},\eta)=\frac{(\eta T_{{\rm obs},i})^k\exp(-\eta T_{{\rm obs},i})}{k!},
\end{equation}
where $\eta$ is the intrinsic burst rate and $T_{{\rm obs},i}$ is the duration of the $i$-th observation. The nondetection probability for the $i$-th observation can be written as
\begin{equation}\label{eq:Pnon}
    P_{{\rm non},i}=1-P(k>0|T_{{\rm obs},i},\eta)\int_{L_{{\rm min},i}}^{\infty}\phi(L|L_{\rm c},\sigma_{\rm L})\mathrm d L,
\end{equation}
where $L_{{\rm min},i}$ is the minimum detectable luminosity at the $i$-th observation. It can be calculated using the minimum detectable flux density $S_{{\rm min},i}$ for the $i$-th observation (as listed in \autoref{tab:smin}) and the luminosity distances $d_{\rm L}(z)$, i.e.,
\begin{equation}
    L_{{\rm min},i}=4\pi d_{\rm L}^2(z) \Delta \nu S_{{\rm min},i}.
\end{equation}
Consequently, the total nondetection probability across multiple independent observations can be calculated as the product of the individual nondetection probabilities, i.e., $P_{\rm non}=\prod_{i}P_{{\rm non},i}$.

By adopting a nondetection probability of $P_{\rm non}=95\%$, we place constraints on the bust rate $\eta$ as a function of the luminosity function parameters $L_{\rm c}$ and $\sigma_{\rm L}$, as shown in \autoref{fig:lf}. The color in each pixel corresponds to the burst rate $\eta$ required for a given $L_{\rm c}$ and $\sigma_{\rm L}$ to remain consistent with our nondetection. For a given detection probability, larger values of $\sigma_{\rm L}$ and smaller values of $L_{\rm c}$ require a higher $\eta$ to produce a sufficient number of bursts. While for higher values of $L_{\rm c}$ and smaller values of $\sigma_{\rm L}$, a lower $\eta$ is required. The values of $\sigma_{\rm L}$ and $L_{\rm c}$ for FRB~121102 \citep{Men+2019} are indicated with a white star. If the potential magnetar in FBOTs emits pulses with a luminosity function similar to that of FRB~121102, a burst rate $\eta$ of $<0.01~{\rm hr^{-1}}$ would be required to be consistent with our nondetection.

Assuming that the potential magnetar in FBOTs emits FRB~121102-like pulses, we can calculate a nondetection probability of $10^{-6}$ by assuming $\eta=3~{\rm h^{-1}}$, $L_{\rm c}=10^{39.9}~{\rm erg~s^{-1}}$ and $\sigma_{\rm L}=0.63$ for FRB~121102 \citep{Men+2019}. This extremely low probability is inconsistent with our observations, suggesting that the central engine in FBOTs is probably not composed of highly energetic magnetar-like objects similar to the one powering FRB 121102. Nevertheless, other scenarios cannot be ruled out: (1) FRB~121102 might be an exceptionally active repeater powered by an extraordinarily active magnetar with extremely high levels of activity \citep{Beloborodov+2017,Palaniswamy+2018,Caleb+2019}. If magnetars in FBOTs are less active (e.g., with a lower $L_{\rm c}$ while $\sigma_{\rm L}$ remains unchanged), a higher burst rate $\eta$ would be necessary to align with the nondetection probability (see \autoref{fig:lf}). (2) The radio bursts from FBOTs could be highly collimated \citep{Katz+2017,Lu+2020,Connor+2020,LiDZ+2024} and not beamed in the direction of Earth. When considering beaming in \autoref{eq:Pnon}, by introducing a beaming fraction less than unity in the second term, a higher burst rate $\eta$ or a higher $L_{\rm c}$ (assuming $\sigma_{\rm L}$ remains unchanged) would be required to account for the nondetection. It is also important to note that beaming can artificially increase the apparent isotropic luminosity. (3) The coherent radio emission may be scattered by the surrounding atmosphere, making it undetectable (see discussion below).

\subsection{Can Radio Bursts Escape the FBOT Ejecta?}
To estimate when the potential radio bursts are able to escape from the FBOT ejecta, we consider the ejecta environment of the burst site. When the free-free optical depth for the ejecta reaches unity, radio waves can escape on a time scale \citep[e.g.,][]{Metzger+2017,Palliyaguru+2021}
\begin{equation}
    t_{\rm es}=4\times f^{2/5}_{\rm ion}\nu_{9}^{-2/5}T_4^{-3/10}M_{\rm 0.1}^{2/5}v_9^{-1}~{\rm yr},
\end{equation}
where $f_{\rm ion}$ is the ionized fraction of the ejecta, $\nu_{\rm 9}$ is the observing frequency in units of  ${10^9}$ Hz, $M_{\rm 0.1}$ represents the ejecta mass $M_{\rm ej}$ in units of $0.1~{\rm M_\odot}$, $T_{\rm 4}$ is the ejecta temperature in units of $10^{4}~{\rm K}$, and $v_{\rm 9}$ is the ejecta velocity in units of $10^{9}~{\rm cm~s^{-1}}$. Taking typical FBOT parameters: ejecta mass of $0.1~{\rm M_\odot}$, $f_{\rm ion}\simeq 0.4$, $v_9=1$, and $T_4=1$, one can estimate that radio burst emission at $1.25~{\rm GHz}$ may break out after $2.6~{\rm yr}$. Since our observations were conducted 4 and 6 yr after the explosions for AT2018cow and CSS161010, repestively, it is unlikely that the non-detection of FRBs is primarily due to absorption by ejecta of FBOTs. 
Other target searches for FRBs from long gamma-ray bursts \citep[e.g.,][]{Palliyaguru+2021,Xu+2025} and super-luminous supernovae \citep[e.g.,][]{Law+2019}, which are also thought to be possibly powered by the magnetar central engines, have similarly yielded nondetections. However, their environments differ from those of FBOTs: the ejecta masses for long gamma-ray bursts and super-luminous supernovae are significantly higher ($\sim 1-10~{\rm M_\odot}$), resulting in a longer time scale $t_{\rm es}\sim7-16~{\rm yr}$ for radio bursts escaping from the ejecta.

In addition to free-free absorption, induced Compton scattering can significantly suppress FRB signals in the dense environments surrounding magnetars, especially at low radio frequencies \citep{Lyubarsky+2016,Metzger+2019,Beloborodov+2020}. This suppression effect is generic to all magnetar-powered FRBs and is not unique to those potentially associated with FBOTs. The Galactic magnetar SGR~1935$+$2154, which produced a millisecond-duration radio burst (FRB~200428) along with a short X-ray flare, serves as an illuminating case study \citep[e.g.,][]{Ioka2020,Yu+2021}. As demonstrated by \cite{Ioka2020}, such high-temperature X-ray burst can generate a dense electron-positron outflow that precedes the FRB, inducing strong induced Compton scattering. FRB photons can only escape this outflow if the emission radius exceeds several tens of the magnetar's radius. These factors suggest that the nondetections in our FRB search may not simply be due to the lack of intrinsic FRB activity or unfavorable geometric beaming. Instead, they may also be a consequence of strong propagation effects in the local magnetar environment.

\section{Conclusions}\label{sec:con}
We conducted a single-pulse search using FAST to look for FRBs from two well-localized nearby FBOTs: AT2018cow and CSS161010. Their central engines were proposed to be newborn energetic magnetars \citep{Prentice+2018,Li+2024}. Despite FAST's high sensitivity and time resolution, we detected no FRB-like bursts. Based on these results we derived the most stringent constraint on the radio fluxes of millisecond transients from FBOTs to be $\sim 10~{\rm mJy}$. Assuming a log-normal luminosity function with parameter similar to those of FRB~121102, we found the burst rate of potential magnetars in FBOTs to be $<0.01~{\rm h^{-1}}$. Magnetar-powered transients, encompassing FBOTs, long gamma-ray bursts, and super-luminous supernovae, have been put forward as potential progenitors of repeating FRBs \citep{Metzger+2017,Law+2019,Men+2019,Palliyaguru+2021,Xu+2025}. Our nondetections place constraints on FRB-generating capabilities of magnetars that originate from FBOTs. It is possible that these newborn magnetars either emit FRBs with significantly lower luminosities or at much lower rates compared to known repeating FRB sources, or their active bursting phase does not coincide with our observation period. Our analysis indicates that the timescale for a FRB signal to escape the explosion ejecta in FBOTs is shorter than that for long gamma-ray bursts and super-luminous supernovae. This suggests FBOTs are promising targets for FRB searches. Detecting an FRB from FBOTs would provide definite evidence of magnetar birth. Therefore, further deep radio follow-up observations of new FBOTs could potentially capture FRB events.

\section*{Acknowledgments}
We are grateful to the anonymous referee for carefully reading and for providing insightful comments.
This work made use of the data from FAST (Five-hundred-meter Aperture Spherical radio Telescope)(\url{https://cstr.cn/31116.02.FAST}). FAST is a Chinese national mega-science facility, operated by National Astronomical Observatories, Chinese Academy of Sciences.
S.J.G. acknowledges support from the National Natural Science Foundation of China (NSFC) under grant No.~123B2045.
X.D.L. acknowledges support from the National Key Research and Development Program of China (2021YFA0718500), the National Natural Science Foundation of China (NSFC) under grant No.~12041301, 12121003 and 12203051. P.Z. acknowledges support from the National Natural Science Foundation of China (NSFC) under grant No.~12273010. 
P.W. acknowledges support from the National Natural Science Foundation of China (NSFC) Programs No.~11988101, 12041303, the CAS Youth Interdisciplinary Team, the Youth Innovation Promotion Association CAS (id.~2021055), and the Cultivation Project for FAST Scientific Payoff and Research Achievement of CAMS-CAS.
The computation was made by using the facilities at the High-Performance Computing Center of Collaborative Innovation Center of Advanced Microstructures (Nanjing University).

\section*{Data Availability}
Original FAST observational data are open source in the FAST Data
Center according to the FAST data one-year protection policy.

\vspace{5mm}
\facilities{Five-hundred-meter Aperture Spherical radio Telescope (FAST).}

\software{\texttt{FRB~Fake} (Author: Leon Houben, \url{https://gitlab.com/houben.ljm/frb-faker}), \texttt{DSPSR} \citep{vanStraten+2011}, 
\texttt{Heimdall} \citep{Barsdell+2024}, 
\texttt{PRESTO} \citep{Ransom+2011}, 
\texttt{PSRCHIVE} \citep{Hotan+2004}, 
\texttt{PulsarX} \citep{PulsarX}, 
\texttt{PyGEDM} \citep{pygedm}, 
\texttt{TransientX} \citep{TransientX} and 
\texttt{YOUR} \citep{Aggarwal+2020}}

\bibliography{ref}{}

\begin{thebibliography}{}
\expandafter\ifx\csname natexlab\endcsname\relax\def\natexlab#1{#1}\fi
\providecommand{\url}[1]{\href{#1}{#1}}
\providecommand{\dodoi}[1]{doi:~\href{http://doi.org/#1}{\nolinkurl{#1}}}
\providecommand{\doeprint}[1]{\href{http://ascl.net/#1}{\nolinkurl{http://ascl.net/#1}}}
\providecommand{\doarXiv}[1]{\href{https://arxiv.org/abs/#1}{\nolinkurl{https://arxiv.org/abs/#1}}}

\bibitem[{{Agarwal} {et~al.}(2020){Agarwal}, {Aggarwal}, {Burke-Spolaor},
  {Lorimer}, \& {Garver-Daniels}}]{FETCH}
{Agarwal}, D., {Aggarwal}, K., {Burke-Spolaor}, S., {Lorimer}, D.~R., \&
  {Garver-Daniels}, N. 2020, \mnras, 497, 1661, \dodoi{10.1093/mnras/staa1856}

\bibitem[{{Aggarwal} {et~al.}(2020){Aggarwal}, {Agarwal}, {Kania}, {Fiore},
  {Thomas}, {Ransom}, {Demorest}, {Wharton}, {Burke-Spolaor}, {Lorimer},
  {Mclaughlin}, \& {Garver-Daniels}}]{Aggarwal+2020}
{Aggarwal}, K., {Agarwal}, D., {Kania}, J., {et~al.} 2020, The Journal of Open
  Source Software, 5, 2750, \dodoi{10.21105/joss.02750}

\bibitem[{{Barsdell} \& {Jameson}(2024)}]{Barsdell+2024}
{Barsdell}, B.~R., \& {Jameson}, A. 2024, {Heimdall: GPU accelerated transient
  detection pipeline for radio astronomy}, Astrophysics Source Code Library,
  record ascl:2407.016

\bibitem[{{Beloborodov}(2017)}]{Beloborodov+2017}
{Beloborodov}, A.~M. 2017, \apjl, 843, L26, \dodoi{10.3847/2041-8213/aa78f3}

\bibitem[{{Beloborodov}(2020)}]{Beloborodov+2020}
---. 2020, \apj, 896, 142, \dodoi{10.3847/1538-4357/ab83eb}

\bibitem[{{Bhardwaj} {et~al.}(2024){Bhardwaj}, {Michilli}, {Kirichenko},
  {Modilim}, {Shin}, {Kaspi}, {Andersen}, {Cassanelli}, {Brar}, {Chatterjee},
  {Cook}, {Dong}, {Fonseca}, {Gaensler}, {Ibik}, {Kaczmarek}, {Lanman},
  {Leung}, {Masui}, {Pandhi}, {Pearlman}, {Petroff}, {Pleunis}, {Prochaska},
  {Rafiei-Ravandi}, {Sand}, {Scholz}, \& {Smith}}]{Bhardwaj+2024}
{Bhardwaj}, M., {Michilli}, D., {Kirichenko}, A.~Y., {et~al.} 2024, \apjl, 971,
  L51, \dodoi{10.3847/2041-8213/ad64d1}

\bibitem[{{Bochenek} {et~al.}(2020){Bochenek}, {Ravi}, {Belov}, {Hallinan},
  {Kocz}, {Kulkarni}, \& {McKenna}}]{Bochenek+2020}
{Bochenek}, C.~D., {Ravi}, V., {Belov}, K.~V., {et~al.} 2020, \nat, 587, 59,
  \dodoi{10.1038/s41586-020-2872-x}

\bibitem[{{Borghese} {et~al.}(2022){Borghese}, {Coti Zelati}, {Israel},
  {Pilia}, {Burgay}, {Trudu}, {Zane}, {Turolla}, {Rea}, {Esposito},
  {Mereghetti}, {Tiengo}, \& {Possenti}}]{Borghese+2022}
{Borghese}, A., {Coti Zelati}, F., {Israel}, G.~L., {et~al.} 2022, \mnras, 516,
  602, \dodoi{10.1093/mnras/stac1314}

\bibitem[{{Burke-Spolaor} {et~al.}(2012){Burke-Spolaor}, {Johnston}, {Bailes},
  {Bates}, {Bhat}, {Burgay}, {Champion}, {D'Amico}, {Keith}, {Kramer}, {Levin},
  {Milia}, {Possenti}, {Stappers}, \& {van Straten}}]{BurkeSpolaor+2012}
{Burke-Spolaor}, S., {Johnston}, S., {Bailes}, M., {et~al.} 2012, \mnras, 423,
  1351, \dodoi{10.1111/j.1365-2966.2012.20998.x}

\bibitem[{{Caleb} {et~al.}(2019){Caleb}, {Stappers}, {Rajwade}, \&
  {Flynn}}]{Caleb+2019}
{Caleb}, M., {Stappers}, B.~W., {Rajwade}, K., \& {Flynn}, C. 2019, \mnras,
  484, 5500, \dodoi{10.1093/mnras/stz386}

\bibitem[{{Cheng} \& {Yu}(2014)}]{Cheng+2014}
{Cheng}, Q., \& {Yu}, Y.-W. 2014, \apjl, 786, L13,
  \dodoi{10.1088/2041-8205/786/2/L13}

\bibitem[{{CHIME/FRB Collaboration} {et~al.}(2020){CHIME/FRB Collaboration},
  {Andersen}, {Bandura}, {Bhardwaj}, {Bij}, {Boyce}, {Boyle}, {Brar},
  {Cassanelli}, {Chawla}, {Chen}, {Cliche}, {Cook}, {Cubranic}, {Curtin},
  {Denman}, {Dobbs}, {Dong}, {Fandino}, {Fonseca}, {Gaensler}, {Giri}, {Good},
  {Halpern}, {Hill}, {Hinshaw}, {H{\"o}fer}, {Josephy}, {Kania}, {Kaspi},
  {Landecker}, {Leung}, {Li}, {Lin}, {Masui}, {McKinven}, {Mena-Parra},
  {Merryfield}, {Meyers}, {Michilli}, {Milutinovic}, {Mirhosseini},
  {M{\"u}nchmeyer}, {Naidu}, {Newburgh}, {Ng}, {Patel}, {Pen},
  {Pinsonneault-Marotte}, {Pleunis}, {Quine}, {Rafiei-Ravandi}, {Rahman},
  {Ransom}, {Renard}, {Sanghavi}, {Scholz}, {Shaw}, {Shin}, {Siegel}, {Singh},
  {Smegal}, {Smith}, {Stairs}, {Tan}, {Tendulkar}, {Tretyakov}, {Vanderlinde},
  {Wang}, {Wulf}, \& {Zwaniga}}]{ChimeFrbJ1935+2020}
{CHIME/FRB Collaboration}, {Andersen}, B.~C., {Bandura}, K.~M., {et~al.} 2020,
  \nat, 587, 54, \dodoi{10.1038/s41586-020-2863-y}

\bibitem[{{CHIME/FRB Collaboration} {et~al.}(2025){CHIME/FRB Collaboration},
  {Amiri}, {Amouyal}, {Andersen}, {Andrew}, {Bandura}, {Bhardwaj}, {Boyle},
  {Brar}, {Cassity}, {Chatterjee}, {Curtin}, {Dobbs}, {Dong}, {Dong}, {Eadie},
  {Eftekhari}, {Fong}, {Fonseca}, {Gaensler}, {Halpern}, {Hessels}, {Hopkins},
  {Ibik}, {Joseph}, {Kaczmarek}, {Kahinga}, {Kaspi}, {Khairy}, {Kilpatrick},
  {Lanman}, {Lazda}, {Leung}, {Main}, {Mas-Ribas}, {Masui}, {Mckinven},
  {Mena-Parra}, {Meyers}, {Michilli}, {Milutinovic}, {Nimmo}, {Noble},
  {Pandhi}, {Shivraj Patil}, {Pearlman}, {Petroff}, {Pleunis}, {Prochaska},
  {Rafiei-Ravandi}, {Rahman}, {Renard}, {Sammons}, {Sand}, {Scholz}, {Shah},
  {Shin}, {Siegel}, {Simha}, {Smith}, {Stairs}, {Vanderlinde}, {Wang}, {Wulf},
  \& {Zegmott}}]{chimefrb+2025}
{CHIME/FRB Collaboration}, {Amiri}, M., {Amouyal}, D., {et~al.} 2025, arXiv
  e-prints, arXiv:2502.11217, \dodoi{10.48550/arXiv.2502.11217}

\bibitem[{{Chrimes} {et~al.}(2024){Chrimes}, {Coppejans}, {Jonker}, {Levan},
  {Groot}, {Mummery}, \& {Stanway}}]{Chrimes+2024}
{Chrimes}, A.~A., {Coppejans}, D.~L., {Jonker}, P.~G., {et~al.} 2024, \aap,
  691, A329, \dodoi{10.1051/0004-6361/202451172}

\bibitem[{{Connor} {et~al.}(2020){Connor}, {Miller}, \&
  {Gardenier}}]{Connor+2020}
{Connor}, L., {Miller}, M.~C., \& {Gardenier}, D.~W. 2020, \mnras, 497, 3076,
  \dodoi{10.1093/mnras/staa2074}

\bibitem[{{Connor} {et~al.}(2025){Connor}, {Ravi}, {Sharma}, {Ocker}, {Faber},
  {Hallinan}, {Harnach}, {Hellbourg}, {Hobbs}, {Hodge}, {Hodges}, {Kosogorov},
  {Lamb}, {Law}, {Rasmussen}, {Sherman}, {Somalwar}, {Weinreb}, {Woody}, \&
  {Konietzka}}]{Connor+2025}
{Connor}, L., {Ravi}, V., {Sharma}, K., {et~al.} 2025, Nature Astronomy,
  \dodoi{10.1038/s41550-025-02566-y}

\bibitem[{{Coppejans} {et~al.}(2020){Coppejans}, {Margutti}, {Terreran},
  {Nayana}, {Coughlin}, {Laskar}, {Alexander}, {Bietenholz}, {Caprioli},
  {Chandra}, {Drout}, {Frederiks}, {Frohmaier}, {Hurley}, {Kochanek},
  {MacLeod}, {Meisner}, {Nugent}, {Ridnaia}, {Sand}, {Svinkin}, {Ward}, {Yang},
  {Baldeschi}, {Chilingarian}, {Dong}, {Esquivia}, {Fong}, {Guidorzi},
  {Lundqvist}, {Milisavljevic}, {Paterson}, {Reichart}, {Shappee}, {Stroh},
  {Valenti}, {Zauderer}, \& {Zhang}}]{Coppejans+2020}
{Coppejans}, D.~L., {Margutti}, R., {Terreran}, G., {et~al.} 2020, \apjl, 895,
  L23, \dodoi{10.3847/2041-8213/ab8cc7}

\bibitem[{{Cordes} \& {Chatterjee}(2019)}]{Cordes+2019}
{Cordes}, J.~M., \& {Chatterjee}, S. 2019, \araa, 57, 417,
  \dodoi{10.1146/annurev-astro-091918-104501}

\bibitem[{{Cordes} {et~al.}(2006){Cordes}, {Freire}, {Lorimer}, {Camilo},
  {Champion}, {Nice}, {Ramachandran}, {Hessels}, {Vlemmings}, {van Leeuwen},
  {Ransom}, {Bhat}, {Arzoumanian}, {McLaughlin}, {Kaspi}, {Kasian}, {Deneva},
  {Reid}, {Chatterjee}, {Han}, {Backer}, {Stairs}, {Deshpande}, \&
  {Faucher-Gigu{\`e}re}}]{Cordes+2006}
{Cordes}, J.~M., {Freire}, P.~C.~C., {Lorimer}, D.~R., {et~al.} 2006, \apj,
  637, 446, \dodoi{10.1086/498335}

\bibitem[{{Dessart} {et~al.}(2007){Dessart}, {Burrows}, {Livne}, \&
  {Ott}}]{Dessart+2007}
{Dessart}, L., {Burrows}, A., {Livne}, E., \& {Ott}, C.~D. 2007, \apj, 669,
  585, \dodoi{10.1086/521701}

\bibitem[{{Drout} {et~al.}(2014){Drout}, {Chornock}, {Soderberg}, {Sanders},
  {McKinnon}, {Rest}, {Foley}, {Milisavljevic}, {Margutti}, {Berger},
  {Calkins}, {Fong}, {Gezari}, {Huber}, {Kankare}, {Kirshner}, {Leibler},
  {Lunnan}, {Mattila}, {Marion}, {Narayan}, {Riess}, {Roth}, {Scolnic},
  {Smartt}, {Tonry}, {Burgett}, {Chambers}, {Hodapp}, {Jedicke}, {Kaiser},
  {Magnier}, {Metcalfe}, {Morgan}, {Price}, \& {Waters}}]{Drout+2014}
{Drout}, M.~R., {Chornock}, R., {Soderberg}, A.~M., {et~al.} 2014, \apj, 794,
  23, \dodoi{10.1088/0004-637X/794/1/23}

\bibitem[{{Duncan} \& {Thompson}(1992)}]{Duncan+1992}
{Duncan}, R.~C., \& {Thompson}, C. 1992, \apjl, 392, L9, \dodoi{10.1086/186413}

\bibitem[{{Gordon} {et~al.}(2023){Gordon}, {Fong}, {Kilpatrick}, {Eftekhari},
  {Leja}, {Prochaska}, {Nugent}, {Bhandari}, {Blanchard}, {Caleb}, {Day},
  {Deller}, {Dong}, {Glowacki}, {Gourdji}, {Mannings}, {Mahoney}, {Marnoch},
  {Miller}, {Paterson}, {Rastinejad}, {Ryder}, {Sadler}, {Scott}, {Sears},
  {Shannon}, {Simha}, {Stappers}, \& {Tejos}}]{Gordon+2023}
{Gordon}, A.~C., {Fong}, W.-f., {Kilpatrick}, C.~D., {et~al.} 2023, \apj, 954,
  80, \dodoi{10.3847/1538-4357/ace5aa}

\bibitem[{{Gourdji} {et~al.}(2019){Gourdji}, {Michilli}, {Spitler}, {Hessels},
  {Seymour}, {Cordes}, \& {Chatterjee}}]{Gourdji+2019}
{Gourdji}, K., {Michilli}, D., {Spitler}, L.~G., {et~al.} 2019, \apjl, 877,
  L19, \dodoi{10.3847/2041-8213/ab1f8a}

\bibitem[{{Ho} {et~al.}(2020){Ho}, {Perley}, {Kulkarni}, {Dong}, {De},
  {Chandra}, {Andreoni}, {Bellm}, {Burdge}, {Coughlin}, {Dekany}, {Feeney},
  {Frederiks}, {Fremling}, {Golkhou}, {Graham}, {Hale}, {Helou}, {Horesh},
  {Kasliwal}, {Laher}, {Masci}, {Miller}, {Porter}, {Ridnaia}, {Rusholme},
  {Shupe}, {Soumagnac}, \& {Svinkin}}]{Ho+2020+at2018lug}
{Ho}, A. Y.~Q., {Perley}, D.~A., {Kulkarni}, S.~R., {et~al.} 2020, \apj, 895,
  49, \dodoi{10.3847/1538-4357/ab8bcf}

\bibitem[{{Ho} {et~al.}(2023){Ho}, {Perley}, {Gal-Yam}, {Lunnan}, {Sollerman},
  {Schulze}, {Das}, {Dobie}, {Yao}, {Fremling}, {Adams}, {Anand}, {Andreoni},
  {Bellm}, {Bruch}, {Burdge}, {Castro-Tirado}, {Dahiwale}, {De}, {Dekany},
  {Drake}, {Duev}, {Graham}, {Helou}, {Kaplan}, {Karambelkar}, {Kasliwal},
  {Kool}, {Kulkarni}, {Mahabal}, {Medford}, {Miller}, {Nordin}, {Ofek},
  {Petitpas}, {Riddle}, {Sharma}, {Smith}, {Stewart}, {Taggart}, {Tartaglia},
  {Tzanidakis}, \& {Winters}}]{Ho+2023}
{Ho}, A. Y.~Q., {Perley}, D.~A., {Gal-Yam}, A., {et~al.} 2023, \apj, 949, 120,
  \dodoi{10.3847/1538-4357/acc533}

\bibitem[{{Hotan} {et~al.}(2004){Hotan}, {van Straten}, \&
  {Manchester}}]{Hotan+2004}
{Hotan}, A.~W., {van Straten}, W., \& {Manchester}, R.~N. 2004, \pasa, 21, 302,
  \dodoi{10.1071/AS04022}

\bibitem[{{Hurley-Walker} {et~al.}(2022){Hurley-Walker}, {Zhang}, {Bahramian},
  {McSweeney}, {O'Doherty}, {Hancock}, {Morgan}, {Anderson}, {Heald}, \&
  {Galvin}}]{Hurley-Walker+2022}
{Hurley-Walker}, N., {Zhang}, X., {Bahramian}, A., {et~al.} 2022, \nat, 601,
  526, \dodoi{10.1038/s41586-021-04272-x}

\bibitem[{{Inserra}(2019)}]{Inserra+2019}
{Inserra}, C. 2019, Nature Astronomy, 3, 697, \dodoi{10.1038/s41550-019-0854-4}

\bibitem[{{Inserra} {et~al.}(2013){Inserra}, {Smartt}, {Jerkstrand}, {Valenti},
  {Fraser}, {Wright}, {Smith}, {Chen}, {Kotak}, {Pastorello}, {Nicholl},
  {Bresolin}, {Kudritzki}, {Benetti}, {Botticella}, {Burgett}, {Chambers},
  {Ergon}, {Flewelling}, {Fynbo}, {Geier}, {Hodapp}, {Howell}, {Huber},
  {Kaiser}, {Leloudas}, {Magill}, {Magnier}, {McCrum}, {Metcalfe}, {Price},
  {Rest}, {Sollerman}, {Sweeney}, {Taddia}, {Taubenberger}, {Tonry},
  {Wainscoat}, {Waters}, \& {Young}}]{Inserra+2013}
{Inserra}, C., {Smartt}, S.~J., {Jerkstrand}, A., {et~al.} 2013, \apj, 770,
  128, \dodoi{10.1088/0004-637X/770/2/128}

\bibitem[{{Ioka}(2020)}]{Ioka2020}
{Ioka}, K. 2020, \apjl, 904, L15, \dodoi{10.3847/2041-8213/abc6a3}

\bibitem[{{Jiang} {et~al.}(2020){Jiang}, {Tang}, {Hou}, {Liu}, {Kr{\v{c}}o},
  {Qian}, {Sun}, {Ching}, {Liu}, {Duan}, {Yue}, {Gan}, {Yao}, {Li}, {Pan},
  {Yu}, {Liu}, {Li}, {Peng}, {Yan}, \& {FAST Collaboration}}]{Jiang+2020}
{Jiang}, P., {Tang}, N.-Y., {Hou}, L.-G., {et~al.} 2020, Research in Astronomy
  and Astrophysics, 20, 064, \dodoi{10.1088/1674-4527/20/5/64}

\bibitem[{{Kasen} \& {Bildsten}(2010)}]{Kasen+2010}
{Kasen}, D., \& {Bildsten}, L. 2010, \apj, 717, 245,
  \dodoi{10.1088/0004-637X/717/1/245}

\bibitem[{{Kashiyama} \& {Quataert}(2015)}]{Kashiyama+2015}
{Kashiyama}, K., \& {Quataert}, E. 2015, \mnras, 451, 2656,
  \dodoi{10.1093/mnras/stv1164}

\bibitem[{{Katz}(2017)}]{Katz+2017}
{Katz}, J.~I. 2017, \mnras, 467, L96, \dodoi{10.1093/mnrasl/slx014}

\bibitem[{{Kulkarni} {et~al.}(2014){Kulkarni}, {Ofek}, {Neill}, {Zheng}, \&
  {Juric}}]{Kulkarni+2014}
{Kulkarni}, S.~R., {Ofek}, E.~O., {Neill}, J.~D., {Zheng}, Z., \& {Juric}, M.
  2014, \apj, 797, 70, \dodoi{10.1088/0004-637X/797/1/70}

\bibitem[{{Law} {et~al.}(2019){Law}, {Omand}, {Kashiyama}, {Murase}, {Bower},
  {Aggarwal}, {Burke-Spolaor}, {Butler}, {Demorest}, {Lazio}, {Linford},
  {Tendulkar}, \& {Rupen}}]{Law+2019}
{Law}, C.~J., {Omand}, C.~M.~B., {Kashiyama}, K., {et~al.} 2019, \apj, 886, 24,
  \dodoi{10.3847/1538-4357/ab4adb}

\bibitem[{{Law} {et~al.}(2024){Law}, {Sharma}, {Ravi}, {Chen}, {Catha},
  {Connor}, {Faber}, {Hallinan}, {Harnach}, {Hellbourg}, {Hobbs}, {Hodge},
  {Hodges}, {Lamb}, {Rasmussen}, {Sherman}, {Shi}, {Simard}, {Squillace},
  {Weinreb}, {Woody}, \& {Yurk}}]{Law2024}
{Law}, C.~J., {Sharma}, K., {Ravi}, V., {et~al.} 2024, \apj, 967, 29,
  \dodoi{10.3847/1538-4357/ad3736}

\bibitem[{{Levin} {et~al.}(2012){Levin}, {Bailes}, {Bates}, {Bhat}, {Burgay},
  {Burke-Spolaor}, {D'Amico}, {Johnston}, {Keith}, {Kramer}, {Milia},
  {Possenti}, {Stappers}, \& {van Straten}}]{Levin+2012}
{Levin}, L., {Bailes}, M., {Bates}, S.~D., {et~al.} 2012, \mnras, 422, 2489,
  \dodoi{10.1111/j.1365-2966.2012.20807.x}

\bibitem[{{Li} {et~al.}(2021){Li}, {Lin}, {Xiong}, {Ge}, {Li}, {Li}, {Lu},
  {Zhang}, {Tuo}, {Nang}, {Zhang}, {Xiao}, {Chen}, {Song}, {Xu}, {Liu}, {Jia},
  {Cao}, {Qu}, {Zhang}, {Gu}, {Liao}, {Zhao}, {Tan}, {Nie}, {Zhao}, {Zheng},
  {Zheng}, {Luo}, {Cai}, {Li}, {Xue}, {Bu}, {Chang}, {Chen}, {Chen}, {Chen},
  {Chen}, {Chen}, {Cui}, {Cui}, {Deng}, {Dong}, {Du}, {Fu}, {Gao}, {Gao},
  {Gao}, {Gu}, {Guan}, {Guo}, {Han}, {Huang}, {Huo}, {Jiang}, {Jiang}, {Jin},
  {Jin}, {Kong}, {Li}, {Li}, {Li}, {Li}, {Li}, {Li}, {Li}, {Liang}, {Liu},
  {Liu}, {Liu}, {Liu}, {Liu}, {Lu}, {Lu}, {Luo}, {Ma}, {Meng}, {Ou}, {Sai},
  {Shang}, {Song}, {Sun}, {Tao}, {Wang}, {Wang}, {Wang}, {Wang}, {Wang}, {Wen},
  {Wu}, {Wu}, {Wu}, {Xiao}, {Xu}, {Yang}, {Yang}, {Yang}, {Yang}, {Yi}, {Yin},
  {You}, {Zhang}, {Zhang}, {Zhang}, {Zhang}, {Zhang}, {Zhang}, {Zhang},
  {Zhang}, {Zhang}, {Zhang}, {Zhang}, {Zhang}, {Zhang}, {Zhang}, {Zhang},
  {Zhang}, {Zhou}, {Zhou}, {Zhu}, {Zhu}, \& {Zhuang}}]{Li+2021NatAs}
{Li}, C.~K., {Lin}, L., {Xiong}, S.~L., {et~al.} 2021, Nature Astronomy, 5,
  378, \dodoi{10.1038/s41550-021-01302-6}

\bibitem[{{Li} \& {Pen}(2024)}]{LiDZ+2024}
{Li}, D., \& {Pen}, U.-L. 2024, \mnras, 531, 2330,
  \dodoi{10.1093/mnras/stae1190}

\bibitem[{{Li} {et~al.}(2018){Li}, {Wang}, {Qian}, {Krco}, {Jiang}, {Yue},
  {Jin}, {Zhu}, {Pan}, {Nan}, \& {Dunning}}]{LiDi+2018}
{Li}, D., {Wang}, P., {Qian}, L., {et~al.} 2018, IEEE Microwave Magazine, 19,
  112, \dodoi{10.1109/MMM.2018.2802178}

\bibitem[{{Li} {et~al.}(2024){Li}, {Zhong}, {Xiao}, {Dai}, {Huang}, \&
  {Sheng}}]{Li+2024}
{Li}, L., {Zhong}, S.-Q., {Xiao}, D., {et~al.} 2024, \apjl, 963, L13,
  \dodoi{10.3847/2041-8213/ad2611}

\bibitem[{{Liu} {et~al.}(2023){Liu}, {Liu}, {Yu}, \& {Zhu}}]{Liu+2023}
{Liu}, J.-F., {Liu}, L.-D., {Yu}, Y.-W., \& {Zhu}, J.-P. 2023, \apj, 946, 35,
  \dodoi{10.3847/1538-4357/acbb04}

\bibitem[{{Liu} {et~al.}(2022){Liu}, {Zhu}, {Liu}, {Yu}, \& {Zhang}}]{Liu+2022}
{Liu}, J.-F., {Zhu}, J.-P., {Liu}, L.-D., {Yu}, Y.-W., \& {Zhang}, B. 2022,
  \apjl, 935, L34, \dodoi{10.3847/2041-8213/ac86d2}

\bibitem[{{Lorimer}(2011)}]{filterbank}
{Lorimer}, D.~R. 2011, {SIGPROC: Pulsar Signal Processing Programs},
  Astrophysics Source Code Library, record ascl:1107.016

\bibitem[{{Lorimer} {et~al.}(2007){Lorimer}, {Bailes}, {McLaughlin},
  {Narkevic}, \& {Crawford}}]{Lorimer+2007}
{Lorimer}, D.~R., {Bailes}, M., {McLaughlin}, M.~A., {Narkevic}, D.~J., \&
  {Crawford}, F. 2007, Science, 318, 777, \dodoi{10.1126/science.1147532}

\bibitem[{{Lorimer} \& {Kramer}(2004)}]{pulsarBook}
{Lorimer}, D.~R., \& {Kramer}, M. 2004, {Handbook of Pulsar Astronomy}, Vol.~4
  (Cambridge University Press)

\bibitem[{{Lorimer} {et~al.}(2024){Lorimer}, {McLaughlin}, \&
  {Bailes}}]{Lorime+2024}
{Lorimer}, D.~R., {McLaughlin}, M.~A., \& {Bailes}, M. 2024, \apss, 369, 59,
  \dodoi{10.1007/s10509-024-04322-6}

\bibitem[{{Lu} {et~al.}(2020){Lu}, {Kumar}, \& {Zhang}}]{Lu+2020}
{Lu}, W., {Kumar}, P., \& {Zhang}, B. 2020, \mnras, 498, 1397,
  \dodoi{10.1093/mnras/staa2450}

\bibitem[{{Lyubarsky}(2014)}]{Lyubarsky+2014}
{Lyubarsky}, Y. 2014, \mnras, 442, L9, \dodoi{10.1093/mnrasl/slu046}

\bibitem[{{Lyubarsky}(2020)}]{Lyubarsky+2020}
---. 2020, \apj, 897, 1, \dodoi{10.3847/1538-4357/ab97b5}

\bibitem[{{Lyubarsky} \& {Ostrovska}(2016)}]{Lyubarsky+2016}
{Lyubarsky}, Y., \& {Ostrovska}, S. 2016, \apj, 818, 74,
  \dodoi{10.3847/0004-637X/818/1/74}

\bibitem[{{Lyutikov}(2021)}]{Lyutikov+2021}
{Lyutikov}, M. 2021, \apj, 922, 166, \dodoi{10.3847/1538-4357/ac1b32}

\bibitem[{{Lyutikov}(2022)}]{Lyytikov+2022}
---. 2022, \mnras, 515, 2293, \dodoi{10.1093/mnras/stac1717}

\bibitem[{{Lyutikov} \& {Toonen}(2019)}]{Lyutikov+2019}
{Lyutikov}, M., \& {Toonen}, S. 2019, \mnras, 487, 5618,
  \dodoi{10.1093/mnras/stz1640}

\bibitem[{{Margalit} {et~al.}(2019){Margalit}, {Berger}, \&
  {Metzger}}]{Margalit+2019}
{Margalit}, B., {Berger}, E., \& {Metzger}, B.~D. 2019, \apj, 886, 110,
  \dodoi{10.3847/1538-4357/ab4c31}

\bibitem[{{McLaughlin} \& {Cordes}(2003)}]{McLaughlin+2003}
{McLaughlin}, M.~A., \& {Cordes}, J.~M. 2003, \apj, 596, 982,
  \dodoi{10.1086/378232}

\bibitem[{{Men} \& {Barr}(2024)}]{TransientX}
{Men}, Y., \& {Barr}, E. 2024, \aap, 683, A183,
  \dodoi{10.1051/0004-6361/202348247}

\bibitem[{{Men} {et~al.}(2023){Men}, {Barr}, {Clark}, {Carli}, \&
  {Desvignes}}]{PulsarX}
{Men}, Y., {Barr}, E., {Clark}, C.~J., {Carli}, E., \& {Desvignes}, G. 2023,
  \aap, 679, A20, \dodoi{10.1051/0004-6361/202347356}

\bibitem[{{Men} {et~al.}(2019){Men}, {Aggarwal}, {Li}, {Palaniswamy},
  {Burke-Spolaor}, {Lee}, {Luo}, {Demorest}, {Tendulkar}, {Agarwal}, {Young},
  \& {Zhang}}]{Men+2019}
{Men}, Y., {Aggarwal}, K., {Li}, Y., {et~al.} 2019, \mnras, 489, 3643,
  \dodoi{10.1093/mnras/stz2386}

\bibitem[{{Mereghetti} {et~al.}(2020){Mereghetti}, {Savchenko}, {Ferrigno},
  {G{\"o}tz}, {Rigoselli}, {Tiengo}, {Bazzano}, {Bozzo}, {Coleiro},
  {Courvoisier}, {Doyle}, {Goldwurm}, {Hanlon}, {Jourdain}, {von Kienlin},
  {Lutovinov}, {Martin-Carrillo}, {Molkov}, {Natalucci}, {Onori}, {Panessa},
  {Rodi}, {Rodriguez}, {S{\'a}nchez-Fern{\'a}ndez}, {Sunyaev}, \&
  {Ubertini}}]{Mereghetti+2020}
{Mereghetti}, S., {Savchenko}, V., {Ferrigno}, C., {et~al.} 2020, \apjl, 898,
  L29, \dodoi{10.3847/2041-8213/aba2cf}

\bibitem[{{Metzger} {et~al.}(2017){Metzger}, {Berger}, \&
  {Margalit}}]{Metzger+2017}
{Metzger}, B.~D., {Berger}, E., \& {Margalit}, B. 2017, \apj, 841, 14,
  \dodoi{10.3847/1538-4357/aa633d}

\bibitem[{{Metzger} {et~al.}(2019){Metzger}, {Margalit}, \&
  {Sironi}}]{Metzger+2019}
{Metzger}, B.~D., {Margalit}, B., \& {Sironi}, L. 2019, \mnras, 485, 4091,
  \dodoi{10.1093/mnras/stz700}

\bibitem[{{Michilli} {et~al.}(2023){Michilli}, {Bhardwaj}, {Brar}, {Gaensler},
  {Kaspi}, {Kirichenko}, {Masui}, {Mckinven}, {Ng}, {Patel}, {Sand}, {Scholz},
  {Shin}, {Siegel}, {Stairs}, {Cassanelli}, {Cook}, {Dobbs}, {Dong}, {Fonseca},
  {Ibik}, {Kaczmarek}, {Leung}, {Pearlman}, {Petroff}, {Pleunis},
  {Rafiei-Ravandi}, {Sanghavi}, {Shaw}, \& {Tendulkar}}]{Michilli+2023}
{Michilli}, D., {Bhardwaj}, M., {Brar}, C., {et~al.} 2023, \apj, 950, 134,
  \dodoi{10.3847/1538-4357/accf89}

\bibitem[{{Mor} {et~al.}(2023){Mor}, {Livne}, \& {Piran}}]{Mor+2023}
{Mor}, R., {Livne}, E., \& {Piran}, T. 2023, \mnras, 518, 623,
  \dodoi{10.1093/mnras/stac2775}

\bibitem[{{Moriya} \& {Eldridge}(2016)}]{Moriya+2016}
{Moriya}, T.~J., \& {Eldridge}, J.~J. 2016, \mnras, 461, 2155,
  \dodoi{10.1093/mnras/stw1471}

\bibitem[{{Nan}(2008)}]{Nan+2008}
{Nan}, R. 2008, in Society of Photo-Optical Instrumentation Engineers (SPIE)
  Conference Series, Vol. 7012, Ground-based and Airborne Telescopes II, ed.
  L.~M. {Stepp} \& R.~{Gilmozzi}, 70121E, \dodoi{10.1117/12.791288}

\bibitem[{{Nan} {et~al.}(2011){Nan}, {Li}, {Jin}, {Wang}, {Zhu}, {Zhu},
  {Zhang}, {Yue}, \& {Qian}}]{Nan+2011}
{Nan}, R., {Li}, D., {Jin}, C., {et~al.} 2011, International Journal of Modern
  Physics D, 20, 989, \dodoi{10.1142/S0218271811019335}

\bibitem[{{Palaniswamy} {et~al.}(2018){Palaniswamy}, {Li}, \&
  {Zhang}}]{Palaniswamy+2018}
{Palaniswamy}, D., {Li}, Y., \& {Zhang}, B. 2018, \apjl, 854, L12,
  \dodoi{10.3847/2041-8213/aaaa63}

\bibitem[{{Palliyaguru} {et~al.}(2021){Palliyaguru}, {Agarwal}, {Golpayegani},
  {Lynch}, {Lorimer}, {Nguyen}, {Corsi}, \& {Burke-Spolaor}}]{Palliyaguru+2021}
{Palliyaguru}, N.~T., {Agarwal}, D., {Golpayegani}, G., {et~al.} 2021, \mnras,
  501, 541, \dodoi{10.1093/mnras/staa3352}

\bibitem[{{Perley} {et~al.}(2021){Perley}, {Ho}, {Yao}, {Fremling}, {Anderson},
  {Schulze}, {Kumar}, {Anupama}, {Barway}, {Bellm}, {Bhalerao}, {Chen}, {Duev},
  {Galbany}, {Graham}, {Gromadzki}, {Guti{\'e}rrez}, {Ihanec}, {Inserra},
  {Kasliwal}, {Kool}, {Kulkarni}, {Laher}, {Masci}, {Neill}, {Nicholl},
  {Pursiainen}, {van Roestel}, {Sharma}, {Sollerman}, {Walters}, \&
  {Wiseman}}]{Perley+2021}
{Perley}, D.~A., {Ho}, A. Y.~Q., {Yao}, Y., {et~al.} 2021, \mnras, 508, 5138,
  \dodoi{10.1093/mnras/stab2785}

\bibitem[{{Petroff} {et~al.}(2019){Petroff}, {Hessels}, \&
  {Lorimer}}]{Petroff+2019R}
{Petroff}, E., {Hessels}, J.~W.~T., \& {Lorimer}, D.~R. 2019, \aapr, 27, 4,
  \dodoi{10.1007/s00159-019-0116-6}

\bibitem[{{Petroff} {et~al.}(2022){Petroff}, {Hessels}, \&
  {Lorimer}}]{Petroff+2022R}
---. 2022, \aapr, 30, 2, \dodoi{10.1007/s00159-022-00139-w}

\bibitem[{{Pietka} {et~al.}(2015){Pietka}, {Fender}, \& {Keane}}]{Pietka+2015}
{Pietka}, M., {Fender}, R.~P., \& {Keane}, E.~F. 2015, \mnras, 446, 3687,
  \dodoi{10.1093/mnras/stu2335}

\bibitem[{{Planck Collaboration}(2020)}]{Planck+2020}
{Planck Collaboration}. 2020, \aap, 641, A6,
  \dodoi{10.1051/0004-6361/201833910}

\bibitem[{{Popov} \& {Postnov}(2010)}]{Popov+2010}
{Popov}, S.~B., \& {Postnov}, K.~A. 2010, in Evolution of Cosmic Objects
  through their Physical Activity, ed. H.~A. {Harutyunian}, A.~M. {Mickaelian},
  \& Y.~{Terzian}, 129--132.
\newblock \doarXiv{0710.2006}

\bibitem[{{Prentice} {et~al.}(2018){Prentice}, {Maguire}, {Smartt}, {Magee},
  {Schady}, {Sim}, {Chen}, {Clark}, {Colin}, {Fulton}, {McBrien}, {O'Neill},
  {Smith}, {Ashall}, {Chambers}, {Denneau}, {Flewelling}, {Heinze}, {Holoien},
  {Huber}, {Kochanek}, {Mazzali}, {Prieto}, {Rest}, {Shappee}, {Stalder},
  {Stanek}, {Stritzinger}, {Thompson}, \& {Tonry}}]{Prentice+2018}
{Prentice}, S.~J., {Maguire}, K., {Smartt}, S.~J., {et~al.} 2018, \apjl, 865,
  L3, \dodoi{10.3847/2041-8213/aadd90}

\bibitem[{{Price} {et~al.}(2021){Price}, {Flynn}, \& {Deller}}]{pygedm}
{Price}, D.~C., {Flynn}, C., \& {Deller}, A. 2021, \pasa, 38, e038,
  \dodoi{10.1017/pasa.2021.33}

\bibitem[{{Pursiainen} {et~al.}(2018){Pursiainen}, {Childress}, {Smith},
  {Prajs}, {Sullivan}, {Davis}, {Foley}, {Asorey}, {Calcino}, {Carollo},
  {Curtin}, {D'Andrea}, {Glazebrook}, {Gutierrez}, {Hinton}, {Hoormann},
  {Inserra}, {Kessler}, {King}, {Kuehn}, {Lewis}, {Lidman}, {Macaulay},
  {M{\"o}ller}, {Nichol}, {Sako}, {Sommer}, {Swann}, {Tucker}, {Uddin},
  {Wiseman}, {Zhang}, {Abbott}, {Abdalla}, {Allam}, {Annis}, {Avila}, {Brooks},
  {Buckley-Geer}, {Burke}, {Carnero Rosell}, {Carrasco Kind}, {Carretero},
  {Castander}, {Cunha}, {Davis}, {De Vicente}, {Diehl}, {Doel}, {Eifler},
  {Flaugher}, {Fosalba}, {Frieman}, {Garc{\'\i}a-Bellido}, {Gruen}, {Gruendl},
  {Gutierrez}, {Hartley}, {Hollowood}, {Honscheid}, {James}, {Jeltema},
  {Kuropatkin}, {Li}, {Lima}, {Maia}, {Martini}, {Menanteau}, {Ogando},
  {Plazas}, {Roodman}, {Sanchez}, {Scarpine}, {Schindler}, {Smith},
  {Soares-Santos}, {Sobreira}, {Suchyta}, {Swanson}, {Tarle}, {Tucker},
  {Walker}, \& {DES Collaboration}}]{Pursiainen+2018}
{Pursiainen}, M., {Childress}, M., {Smith}, M., {et~al.} 2018, \mnras, 481,
  894, \dodoi{10.1093/mnras/sty2309}

\bibitem[{{Ransom}(2011)}]{Ransom+2011}
{Ransom}, S. 2011, {PRESTO: PulsaR Exploration and Search TOolkit},
  Astrophysics Source Code Library, record ascl:1107.017

\bibitem[{{Ridnaia} {et~al.}(2021){Ridnaia}, {Svinkin}, {Frederiks}, {Bykov},
  {Popov}, {Aptekar}, {Golenetskii}, {Lysenko}, {Tsvetkova}, {Ulanov}, \&
  {Cline}}]{Ridnaia+2021NatAs}
{Ridnaia}, A., {Svinkin}, D., {Frederiks}, D., {et~al.} 2021, Nature Astronomy,
  5, 372, \dodoi{10.1038/s41550-020-01265-0}

\bibitem[{{Shannon} {et~al.}(2025){Shannon}, {Bannister}, {Bera}, {Bhandari},
  {Day}, {Deller}, {Dial}, {Dobie}, {Ekers}, {Fong}, {Glowacki}, {Gordon},
  {Gourdji}, {Jaini}, {James}, {Kumar}, {Mahony}, {Marnoch}, {Muller},
  {Prochaska}, {Qiu}, {Ryder}, {Sadler}, {Scott}, {Tejos}, {Uttarkar}, \&
  {Wang}}]{Shannon+2025}
{Shannon}, R.~M., {Bannister}, K.~W., {Bera}, A., {et~al.} 2025, \pasa, 42,
  e036, \dodoi{10.1017/pasa.2025.8}

\bibitem[{{Sharma} {et~al.}(2024){Sharma}, {Ravi}, {Connor}, {Law}, {Ocker},
  {Sherman}, {Kosogorov}, {Faber}, {Hallinan}, {Harnach}, {Hellbourg}, {Hobbs},
  {Hodge}, {Hodges}, {Lamb}, {Rasmussen}, {Somalwar}, {Weinreb}, {Woody},
  {Leja}, {Anand}, {Das}, {Qin}, {Rose}, {Dong}, {Miller}, \&
  {Yao}}]{Sharma+2024}
{Sharma}, K., {Ravi}, V., {Connor}, L., {et~al.} 2024, \nat, 635, 61,
  \dodoi{10.1038/s41586-024-08074-9}

\bibitem[{{Sherman} {et~al.}(2024){Sherman}, {Connor}, {Ravi}, {Law}, {Chen},
  {Catha}, {Faber}, {Hallinan}, {Harnach}, {Hellbourg}, {Hobbs}, {Hodge},
  {Hodges}, {Lamb}, {Rasmussen}, {Sharma}, {Shi}, {Simard}, {Somalwar},
  {Squillace}, {Weinreb}, {Woody}, {Yadlapalli}, \& {The Deep Synoptic Array
  team}}]{Sherman+2024}
{Sherman}, M.~B., {Connor}, L., {Ravi}, V., {et~al.} 2024, \apj, 964, 131,
  \dodoi{10.3847/1538-4357/ad275e}

\bibitem[{{Tauris} {et~al.}(2013){Tauris}, {Langer}, {Moriya}, {Podsiadlowski},
  {Yoon}, \& {Blinnikov}}]{Tauris+2013}
{Tauris}, T.~M., {Langer}, N., {Moriya}, T.~J., {et~al.} 2013, \apjl, 778, L23,
  \dodoi{10.1088/2041-8205/778/2/L23}

\bibitem[{{Tauris} {et~al.}(2015){Tauris}, {Langer}, \&
  {Podsiadlowski}}]{Tauris+2015}
{Tauris}, T.~M., {Langer}, N., \& {Podsiadlowski}, P. 2015, \mnras, 451, 2123,
  \dodoi{10.1093/mnras/stv990}

\bibitem[{{Tauris} {et~al.}(2017){Tauris}, {Kramer}, {Freire}, {Wex}, {Janka},
  {Langer}, {Podsiadlowski}, {Bozzo}, {Chaty}, {Kruckow}, {van den Heuvel},
  {Antoniadis}, {Breton}, \& {Champion}}]{Tauris+2017}
{Tauris}, T.~M., {Kramer}, M., {Freire}, P.~C.~C., {et~al.} 2017, \apj, 846,
  170, \dodoi{10.3847/1538-4357/aa7e89}

\bibitem[{{Tavani} {et~al.}(2021){Tavani}, {Casentini}, {Ursi}, {Verrecchia},
  {Addis}, {Antonelli}, {Argan}, {Barbiellini}, {Baroncelli}, {Bernardi},
  {Bianchi}, {Bulgarelli}, {Caraveo}, {Cardillo}, {Cattaneo}, {Chen}, {Costa},
  {Del Monte}, {Di Cocco}, {Di Persio}, {Donnarumma}, {Evangelista}, {Feroci},
  {Ferrari}, {Fioretti}, {Fuschino}, {Galli}, {Gianotti}, {Giuliani},
  {Labanti}, {Lazzarotto}, {Lipari}, {Longo}, {Lucarelli}, {Magro},
  {Marisaldi}, {Mereghetti}, {Morelli}, {Morselli}, {Naldi}, {Pacciani},
  {Parmiggiani}, {Paoletti}, {Pellizzoni}, {Perri}, {Perotti}, {Piano},
  {Picozza}, {Pilia}, {Pittori}, {Puccetti}, {Pupillo}, {Rapisarda},
  {Rappoldi}, {Rubini}, {Setti}, {Soffitta}, {Trifoglio}, {Trois},
  {Vercellone}, {Vittorini}, {Giommi}, \& {D'Amico}}]{Tavani+2021NatAs}
{Tavani}, M., {Casentini}, C., {Ursi}, A., {et~al.} 2021, Nature Astronomy, 5,
  401, \dodoi{10.1038/s41550-020-01276-x}

\bibitem[{{Thompson}(2023)}]{Thompson+2023}
{Thompson}, C. 2023, \mnras, 519, 497, \dodoi{10.1093/mnras/stac3565}

\bibitem[{{Thornton} {et~al.}(2013){Thornton}, {Stappers}, {Bailes},
  {Barsdell}, {Bates}, {Bhat}, {Burgay}, {Burke-Spolaor}, {Champion}, {Coster},
  {D'Amico}, {Jameson}, {Johnston}, {Keith}, {Kramer}, {Levin}, {Milia}, {Ng},
  {Possenti}, \& {van Straten}}]{Thornton+2013}
{Thornton}, D., {Stappers}, B., {Bailes}, M., {et~al.} 2013, Science, 341, 53,
  \dodoi{10.1126/science.1236789}

\bibitem[{{van Straten} \& {Bailes}(2011)}]{vanStraten+2011}
{van Straten}, W., \& {Bailes}, M. 2011, \pasa, 28, 1, \dodoi{10.1071/AS10021}

\bibitem[{{van Straten} {et~al.}(2010){van Straten}, {Manchester}, {Johnston},
  \& {Reynolds}}]{psrfits}
{van Straten}, W., {Manchester}, R.~N., {Johnston}, S., \& {Reynolds}, J.~E.
  2010, \pasa, 27, 104, \dodoi{10.1071/AS09084}

\bibitem[{{Wadiasingh} \& {Timokhin}(2019)}]{Wadiasingh+2019}
{Wadiasingh}, Z., \& {Timokhin}, A. 2019, \apj, 879, 4,
  \dodoi{10.3847/1538-4357/ab2240}

\bibitem[{{Wang} \& {Liu}(2020)}]{Wangbo+2020}
{Wang}, B., \& {Liu}, D. 2020, Research in Astronomy and Astrophysics, 20, 135,
  \dodoi{10.1088/1674-4527/20/9/135}

\bibitem[{{Wang} {et~al.}(2025){Wang}, {Gao}, \& {Fan}}]{Wang+2025}
{Wang}, Y.-Y., {Gao}, S.-J., \& {Fan}, Y.-Z. 2025, \apj, 981, 9,
  \dodoi{10.3847/1538-4357/adade8}

\bibitem[{{Xu} {et~al.}(2025){Xu}, {Anderson}, {Tian}, {Meyers}, {Tingay},
  {Huang}, {Wang}, {Venville}, {Lee}, {Rowlinson}, {Hancock}, {Williams}, \&
  {Sokolowski}}]{Xu+2025}
{Xu}, F., {Anderson}, G.~E., {Tian}, J., {et~al.} 2025, \apj, 982, 32,
  \dodoi{10.3847/1538-4357/adb71e}

\bibitem[{{Xu} {et~al.}(2023){Xu}, {Feng}, {Li}, {Wang}, {Zhang}, {Xie},
  {Chen}, {Wang}, {Kang}, {Hu}, {Zheng}, {Tsai}, {Chen}, \& {Zhou}}]{Xu+2023}
{Xu}, J., {Feng}, Y., {Li}, D., {et~al.} 2023, Universe, 9, 330,
  \dodoi{10.3390/universe9070330}

\bibitem[{{Yao} {et~al.}(2017){Yao}, {Manchester}, \& {Wang}}]{YMW16}
{Yao}, J.~M., {Manchester}, R.~N., \& {Wang}, N. 2017, \apj, 835, 29,
  \dodoi{10.3847/1538-4357/835/1/29}

\bibitem[{{Yao} {et~al.}(2022){Yao}, {Ho}, {Medvedev}, {Nayana}, {Perley},
  {Kulkarni}, {Chandra}, {Sazonov}, {Gilfanov}, {Khorunzhev}, {Khatami}, \&
  {Sunyaev}}]{Yao+2021}
{Yao}, Y., {Ho}, A. Y.~Q., {Medvedev}, P., {et~al.} 2022, \apj, 934, 104,
  \dodoi{10.3847/1538-4357/ac7a41}

\bibitem[{{Yu} {et~al.}(2019){Yu}, {Chen}, \& {Wang}}]{Yu+2019}
{Yu}, Y.-W., {Chen}, A., \& {Wang}, B. 2019, \apjl, 870, L23,
  \dodoi{10.3847/2041-8213/aaf960}

\bibitem[{{Yu} {et~al.}(2021){Yu}, {Zou}, {Dai}, \& {Yu}}]{Yu+2021}
{Yu}, Y.-W., {Zou}, Y.-C., {Dai}, Z.-G., \& {Yu}, W.-F. 2021, \mnras, 500,
  2704, \dodoi{10.1093/mnras/staa3374}

\bibitem[{{Zenati} {et~al.}(2019){Zenati}, {Perets}, \& {Toonen}}]{Zenati+2019}
{Zenati}, Y., {Perets}, H.~B., \& {Toonen}, S. 2019, \mnras, 486, 1805,
  \dodoi{10.1093/mnras/stz316}

\bibitem[{{Zhang}(2020)}]{Zhang+2020}
{Zhang}, B. 2020, \nat, 587, 45, \dodoi{10.1038/s41586-020-2828-1}

\bibitem[{{Zhang}(2022)}]{Zhang+2022}
---. 2022, \apj, 925, 53, \dodoi{10.3847/1538-4357/ac3979}

\bibitem[{{Zhang}(2023)}]{Zhang+2023R}
---. 2023, Reviews of Modern Physics, 95, 035005,
  \dodoi{10.1103/RevModPhys.95.035005}

\end{thebibliography}
\bibliographystyle{aasjournal}
\end{document}